\documentclass[a4paper]{article}

\usepackage{INTERSPEECH2019}
\usepackage{cite}
\usepackage{balance}
\usepackage{enumitem}
\usepackage{multirow}
\usepackage[stable]{footmisc}
\def\T{{\mathsf T}}
\def\F{{\mathrm F}}

\title{WHAM!: Extending Speech Separation to Noisy Environments}
\name{Gordon Wichern$^1$, Joe Antognini$^2$, Michael Flynn$^2$, Licheng Richard Zhu$^2$,\\ Emmett McQuinn$^2$, Dwight Crow$^2$, Ethan Manilow$^{1}$, Jonathan Le Roux$^1$}
\address{
  $^1$Mitsubishi Electric Research Laboratories (MERL), Cambridge, MA, USA\\
  $^2$Whisper.ai, San Francisco, CA, USA}
\email{\{wichern,leroux\}@merl.com, \{joe,flynn,richard,emmett,dwight\}@whisper.ai}

\begin{document}

\maketitle
\setlength{\abovedisplayskip}{4pt}
\setlength{\belowdisplayskip}{4pt}
\begin{abstract}
  Recent progress in separating the speech signals from multiple overlapping speakers using a single audio channel has brought us closer to solving the cocktail party problem.  However, most studies in this area use a constrained problem setup, comparing performance when speakers overlap almost completely, at artificially low sampling rates, and with no external background noise. In this paper, we strive to move the field towards more realistic and challenging scenarios. To that end, we created the WSJ0 Hipster Ambient Mixtures (WHAM!) dataset, consisting of two speaker mixtures from the wsj0-2mix dataset combined with real ambient noise samples. The samples were collected in coffee shops, restaurants, and bars in the San Francisco Bay Area, and are made publicly available.  We benchmark various speech separation architectures and objective functions to evaluate their robustness to noise.  While separation performance decreases as a result of noise, we still observe substantial gains relative to the noisy signals for most approaches.
\end{abstract}

\section{Introduction}
The problems of speaker-independent monaural speech enhancement (separating speech from background noise) and speech separation (separating multiple overlapping speech signals) have progressed greatly with modern deep learning-based techniques \cite{Weninger2014RNN, Erdogan2015, Hershey2016, williamson2016complex, wang2018supervised, kolbaek2017joint, ephrat2018looking, leroux2019phasebook, Luo2018TasNet09arXiv}.   While high performing enhancement and separation systems share many common techniques, each problem has unique attributes which require specialized solutions.  In enhancement, the typically unstructured background noise may not require accurate reconstruction, but this lack of structure can corrupt the enhanced speech signal in unpredictable ways, for example by significantly degrading the phase information.  When  estimating a time-frequency (T-F) mask that modifies the mixture signal magnitude and uses the noisy mixture phase for resynthesis, the phase-sensitive mask~\cite{Erdogan2015} can help compensate for these noisy phase errors.   

However, in speech separation, both the target and interference signals are highly structured speech requiring accurate reconstruction.  Furthermore, because all outputs are speech signals, we must solve the permutation problem stemming from the fact that the correspondence between the algorithm outputs and the true sources is unknown~\cite{Hershey2016}.  Deep clustering~\cite{Hershey2016, Isik2016, Wang2018ICASSP04Alternative} and permutation-free mask inference~\cite{Hershey2016, Kolbaek2017} are two common approaches for solving the speaker separation problem.  Once permutation is solved, separation may be in some sense easier than enhancement, because networks can better detect patterns in the highly structured speech signals as opposed to unstructured noise.  This has brought forth a novel class of network architectures and objective functions benefiting from some type of phase processing, either implicitly by directly optimizing the time domain waveform~\cite{Wang2018Interspeech09, Luo2018TasNet09arXiv, shi2019furcanext}, or explicitly via phase estimation algorithms~\cite{Wang2018Interspeech09, leroux2019phasebook, wang2018deep}.  Many of these techniques have surpassed the performance of some noisy phase oracle T-F masks~\cite{Luo2018TasNet09arXiv, wang2018deep, shi2019furcanext, Wichern2018IWAENC09} on the benchmark wsj0-2mix dataset~\cite{Hershey2016}.

While the wsj0-2mix dataset has undoubtedly helped to rapidly advance the field of deep learning-based speech separation, it also lacks a certain amount of realism.  Built using utterances from the well-known WSJ0 corpus \cite{garofolo1993csr}, it consists of instantaneous mixtures of two or three simultaneous speakers, without any background noise.  Furthermore, most results reported in the literature use the so-called \emph{min} version of the dataset, which truncates all utterances in a mixture to the length of the shortest utterance; systems are thus trained and evaluated only on near-fully overlapped speech, not on more realistic diarization type scenarios.  Also, to reduce processing and memory consumption, results are typically reported using data downsampled to 8 kHz, ignoring a large part of the speech spectrum. %
To the best of our knowledge, the robustness of speech separation algorithms in noise was only considered in~\cite{kolbaek2017joint, drude2019integration}, but the types and amount of noise
were somewhat limited. 

To help facilitate development and evaluation of speech separation in more realistic scenarios, we introduce the WSJ0 Hipster Ambient Mixtures (WHAM!) dataset, which pairs each two-speaker utterance in the wsj0-2mix dataset with a unique noise background scene, recorded with a binaural microphone in non-stationary ambient environments such as coffee shops, restaurants, and bars.  WHAM! is made publicly available and attempts to maintain parity with the wsj0-2mix dataset so that researchers can easily evaluate the robustness of speech separation algorithms against noise. Additionally, the WHAM! dataset can be used for training and evaluating speech enhancement algorithms.  The initial version of WHAM! considers a single-channel, non-reverberant setup, but extensions to stereo and reverberant conditions are currently under investigation.

In this paper, we carry out a series of initial experiments with the WHAM! dataset for both enhancement and separation.  For enhancement, we evaluate T-F masking approaches based on BLSTM networks trained via the phase-sensitive approximation (PSA) objective~\cite{Erdogan2015}.  We evaluate enhancement performance both in the usual single-speaker case and when removing noise from two overlapping speakers.  For separation, we focus mainly on the chimera++ architecture~\cite{Wang2018ICASSP04Alternative} and evaluate variations of the deep clustering head for simultaneous separation and noise removal.  We report similar objective separation performance when jointly enhancing and separating, and when first running an enhancement algorithm on the two overlapping speakers followed by a separate separation network operating on the enhanced signals.  We also present a subset of benchmark results using various network architectures from the literature on both the enhancement and noisy separation tasks.

\section{WHAM! dataset\footnote{Available at: http://wham.whisper.ai}}
The wsj0-2mix dataset~\cite{Hershey2016} is composed of two-speaker mixtures from the Wall Street Journal (WSJ0) corpus, and scripts for creating this dataset are publicly available.  The mixtures are created by applying randomly selected gains in order to achieve relative levels between 0 and 5 dB between the two speech signals prior to mixing in the time domain.  The dataset contains 20,000, 5,000 and 3,000 instantaneous two-speaker mixtures in its 30~h training, 10~h validation, and 5~h test sets, respectively.
The training and validation sets share common speakers, but the test set speakers are different.  There are four variations of the wsj0-2mix dataset, a \emph{min} version where the longer of the two signals is truncated, and a \emph{max} version where silence is appended to the shorter signal, both available at 16 kHz and 8 kHz sampling rates.  A three-speaker version of wsj0-mix also exists. We have not yet created a corresponding noisy version, but an extension of the approach described in the rest of this section to the three-speaker case is straightforward.

\begin{figure}[tb]
	\centering
		\includegraphics[width=.48\columnwidth]{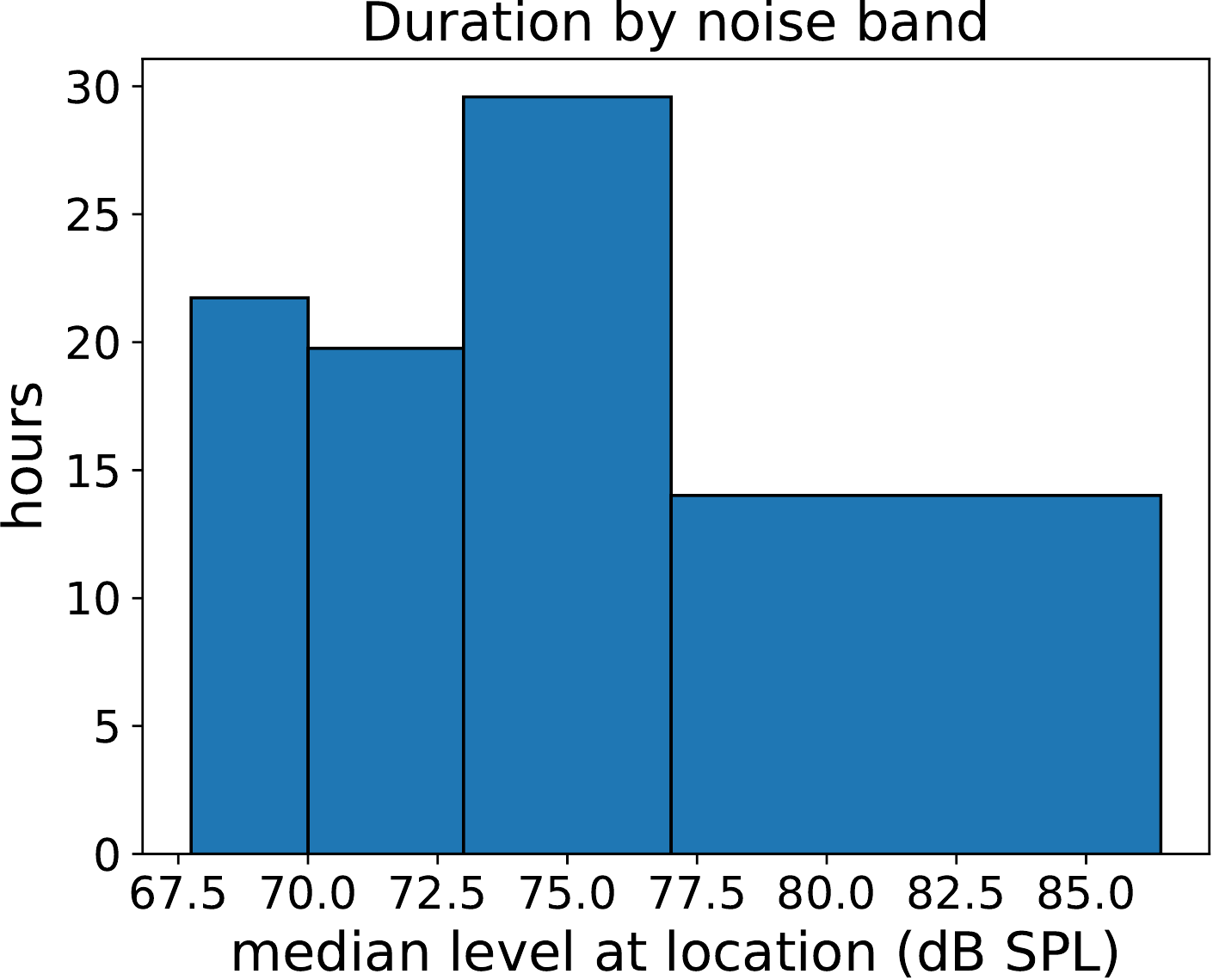}
		\includegraphics[width=.48\columnwidth]{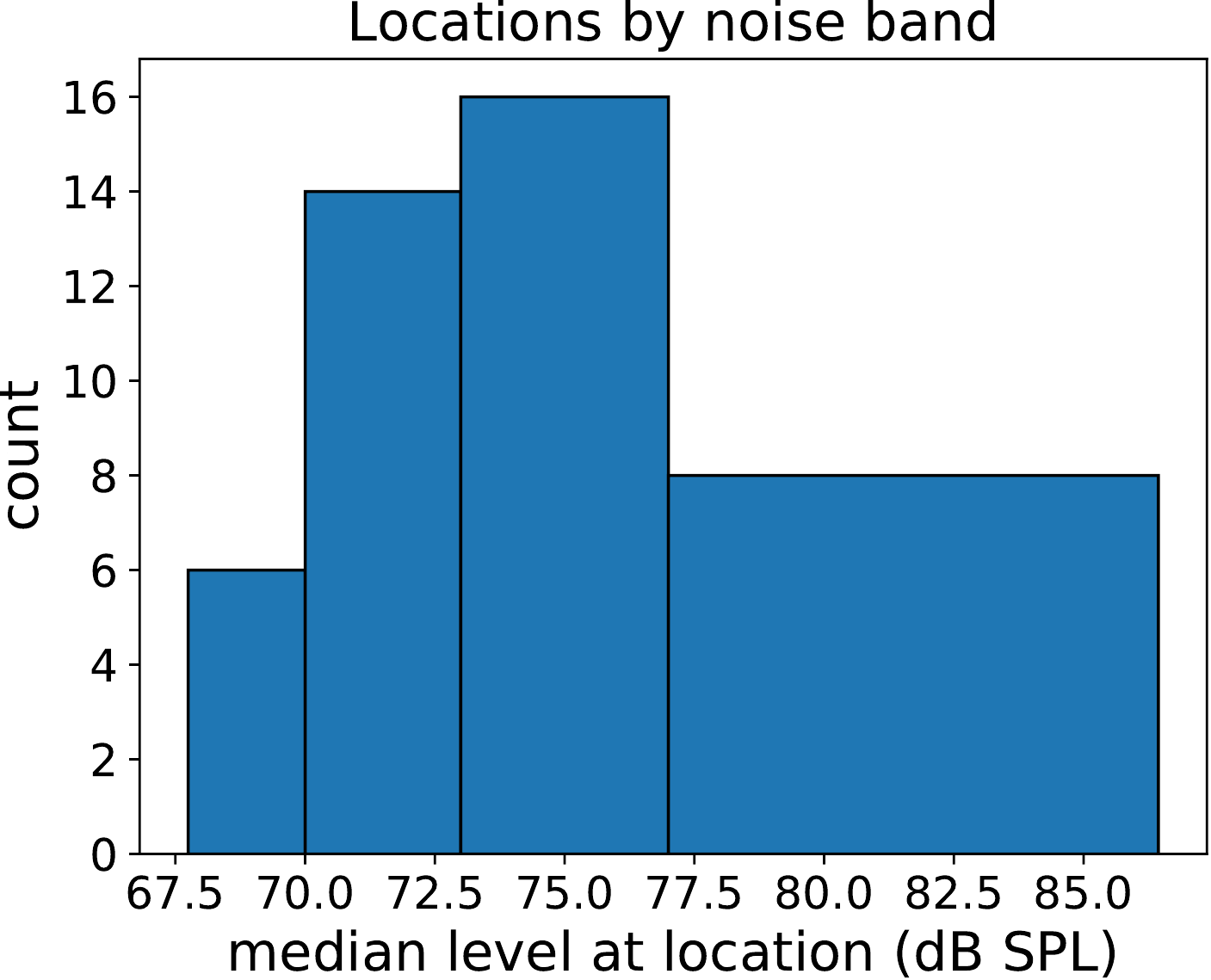}\vspace{-.15cm}
	\caption{Histograms of duration and unique locations where background noise was recorded.}
	\label{fig:dataset_hists}\vspace{-.1cm}
\end{figure}

Our background noise dataset was recorded in urban environments %
such as coffee shops, restaurants, bars, office buildings, parks, etc, in the San Francisco Bay Area.  Audio was recorded using an Apogee Sennheiser binaural microphone connected to a smartphone, where the microphone was mounted on a tripod typically sitting on a table with heights varying between 1.0-1.5 m, and an inter-microphone distance between 15-17 cm.  Pre-amp gain was set to a constant calibrated value prior to each recording.  While the audio is captured at a sampling rate of 48 kHz, we downsample to 16 kHz and 8 kHz to maintain parity with the original wsj0-2mix dataset.  We also only evaluate single-channel approaches in this work, but make the stereo recordings available for consideration in subsequent work. Figure~\ref{fig:dataset_hists} shows sound pressure level (SPL) histograms over all captured ambient recordings, which in raw form consisted of close to 80 hours of audio recorded at 44 different locations (often each location was visited multiple times on different days).  All recording locations were first partitioned into the four bins shown in Fig.~\ref{fig:dataset_hists} based on SPL, roughly corresponding to very quiet, quiet, normal, and loud locations.  Exact bin spacing was chosen such that each bin contained at least six unique locations, and at least 12 hours of audio.  We then assigned all recordings from a given location to either the training, validation, or test split, such that each split contained recordings from at least two unique locations in each bin from Fig.~\ref{fig:dataset_hists}, and the durations were roughly proportional to the 30h/10h/5h training/validation/test sets of the original wsj0-2mix.

\begin{figure}[tb]
	\centering
		\includegraphics[width=.70\columnwidth]{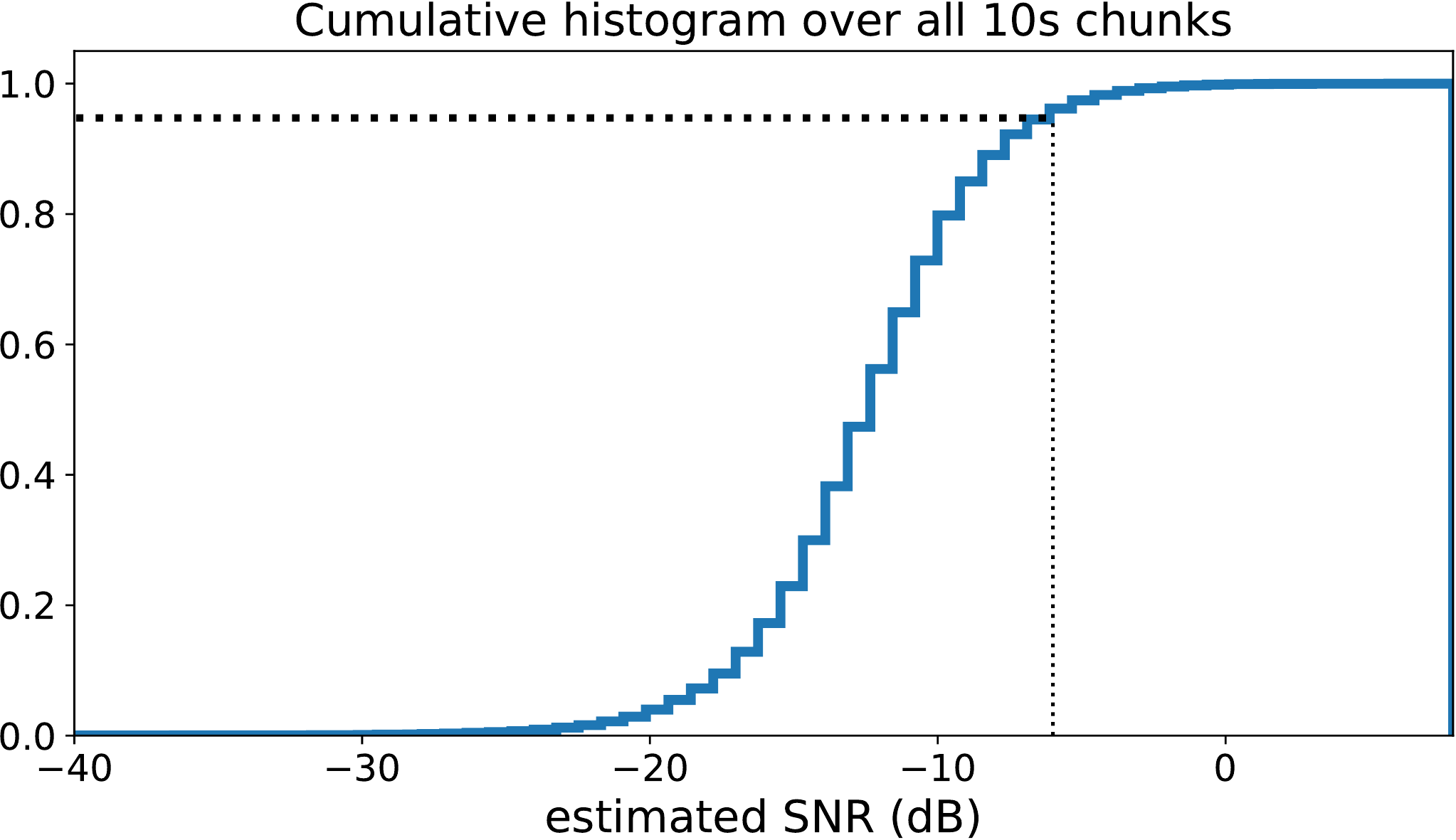}
		\vspace{-.25cm}
	\caption{Estimated speech SNR of all background recordings, with -6 dB threshold indicated.}
	\label{fig:est_snr_cdf}\vspace{-.7cm}
\end{figure}

Because the noise is to be mixed with clean speech to train enhancement and separation models, it is critical that high SNR, intelligible speech be removed from the ambient noise corpus.  To accomplish this, we used an inverted approach to the one used to remove overly noisy speech when creating AVSpeech in~\cite{ephrat2018looking}.  We first process the ambient recordings with the commercially available \emph{iZotope RX 7 Dialogue Isolate} tool to obtain an estimate of any foreground speech.  We then compute the SNR between the estimated foreground speech and the residual for each 10 second chunk of audio.  We only include noise clips if the estimated SNR is less than -6 dB, which eliminated approximately 5\%  of the available data, as shown in Fig.~\ref{fig:est_snr_cdf}.


To maintain parity with wsj0-2mix, we enforce the same relative levels between the two speakers.  Noise is mixed in by first sampling a random SNR value from a uniform distribution between -6 and +3 dB.  We then apply a gain to the first (louder) speaker such that the SNR between the first speaker and the noise is equal to the randomly sampled value.  The SNR range was chosen by recording conversations in some of the same environments in which the ambient noise was collected, and estimating the relative speech and noise levels.  We also examined whether the SNR varied as the level of ambient background noise increased.  We found that the speakers spoke louder and/or moved closer in loud environments, but the SNR-range remained relatively consistent, although many other properties of the speech signal changed due to the Lombard effect~\cite{lu2008speech}.  Note that here we compute SNR using loudness units full-scale (LUFS)~\cite{grimm2010toward} to obtain a more perceptually meaningful scaling and also to remove silent regions from the SNR computation.  The same gain is then applied to the second speaker.  The noise file to use for a given utterance is randomly sampled as follows: (1) sample one of the four noise bands from Fig.~\ref{fig:dataset_hists} uniformly, (2) sample a noise file proportionally to its length, and (3) sample a random portion of the file of appropriate length for the wsj0-2mix max utterance, randomly adding up to two seconds noise before and after the utterance, i.e., up to four seconds total.  We also create a min version of WHAM! by removing any leading and trailing noise and truncating to the length of the shorter of the two speakers.  Scripts for creating WHAM! from wsj0-2mix and the noise clips corresponding to each utterance are publicly available under a Creative Commons license. 

\vspace{-0.2cm}
\section{Speech separation objective functions} \vspace{-0.1cm}
\label{sec:objectives}
Let $X \in \mathbb{C}^{F\times T}$ be the complex spectrogram of a mixture of $C$ sources $S_c \in \mathbb{C}^{F\times T}$ for $c=1,\dots,C$. For simplicity, we focus here mainly on methods that attempt to estimate a real-valued mask for each source $\hat{M}_c \in \mathbb{R}^{F\times T}$ by minimizing the truncated phase sensitive approximation (tPSA) objective \cite{Erdogan2015} in a permutation-free manner \cite{Hershey2016, Isik2016, Kolbaek2017}:
\begin{multline}
	\mathcal{L}_{\text{tPSA}} = \min_{\pi \in \mathcal{P}} \sum_{c} \Big\| \hat{M}_{\pi(c)} \odot |X| 
	\\ - \operatorname{T}_{0}^{|X|}\left(|S_c| \odot \cos(\angle S_c - \angle X)\right) \Big\|_1, \label{eq:L_MI_PSA}
\end{multline}
where $\mathcal{P}$ is the set of all possible permutations over the set of sources $\{1,\dots, C\}$, $\odot$ denotes element-wise product, $\angle S_c$ is the true phase of source c, $\angle X$ is the mixture phase, and $\operatorname{T}_{0}^{|X|}(x)= \min(\max(x,0),|X|)$ is a truncation function ensuring the target can be reached with a sigmoid activation function.  For enhancement, we typically are not interested in including the reconstruction error for the background noise, and the sum term in \eqref{eq:L_MI_PSA} is removed since there is only a single target signal.  Similarly, for noisy separation, the noise is not included in the set of sources over which the loss in \eqref{eq:L_MI_PSA} is computed.

For speech separation, mask estimation can be further improved by incorporating a deep clustering regularization term into the loss function as in the chimera++ architecture~\cite{Wang2018ICASSP04Alternative}, i.e.,
\begin{align} \label{eq:chimera}
	\mathcal{L}_{\text{chi}^{++}_\alpha}=\alpha \mathcal{L}_{\text{DC}}+(1-\alpha)\mathcal{L}_{\text{tPSA}},
\end{align}
where the weight $\alpha$ is typically set to a high value, e.g., 0.975.  The deep clustering loss $\mathcal{L}_{\text{DC}}$ in~\eqref{eq:chimera} can take multiple forms as proposed in~\cite{Wang2018ICASSP04Alternative}, but here we focus on the classic and whitened k-means variations of the objective, i.e., 
\begin{align}\label{eq:dc_classic}
\mathcal{L}_{\text{DC},\text{C}}&\!=\!\|V V^{\T}-YY^{\T}\|_{\F}^2,\\
\label{eq:dc_white}
    \mathcal{L}_{\text{DC},\text{W}}&\! =\!\|V(V^{\T}V)^{-\frac{1}{2}} \!-\!  Y(Y^{\T}Y)^{-1}Y^{\T} V (V^{\T}V)^{-\frac{1}{2}} \|_{\F}^2, \!
\end{align}
where $V\in \mathbb{R}^{TF\times D}$ is an embedding matrix consisting of vertically stacked embedding vectors, and $Y\in \mathbb{R}^{TF\times C}$ is a label matrix consisting of vertically stacked one-hot label vectors representing which of the $c$ sources in a mixture dominates at each T-F bin.  We also discount the influence of low-energy T-F bins by applying magnitude ratio weights~\cite{Wang2018ICASSP04Alternative} to both $V$ and $Y$.  When extending the deep clustering loss to noisy speech separation, we have several options in how we treat the noise source. Our default approach, which is also the most straightforward, is to treat the noise signal like an additional speech signal and use~\eqref{eq:dc_classic} or~\eqref{eq:dc_white}.  An alternative approach is to only include the speech sources in $\mathcal{L}_{\text{DC}}$, and apply a weight of 0 to all T-F bins without speech. Yet another possibility is to consider a \emph{noise-aware deep clustering} objective function that attempts to push embeddings of the noise-dominated T-F bins far from the speech-dominated bins, without enforcing the noise-dominated bins to be close to one another (pairs of speech-dominated T-F bins are handled as usual, with embeddings of bins dominated by the same speaker pushed to be close to each other and far from those dominated by other speakers).  This can be achieved by subtracting the value of $\mathcal{L}_{\text{DC},\text{C}}$ for the noise-dominated bins from the value of $\mathcal{L}_{\text{DC},\text{C}}$ for all T-F bins, i.e.,
\begin{align}\label{eq:dc_noisy}
\mathcal{L}_{\text{DC},\text{N}}=\|V V^{\T}-YY^{\T}\|_{\F}^2 - \|V_n V_n^{\T}-Y_nY_n^{\T}\|_{\F}^2
\end{align}
where $V_n$ and $Y_n$ denote the embedding and label matrix restricted to the noise-dominated T-F bins.  The final approach we consider for speech separation in noise uses two separate networks connected in series: (1) an enhancement network trained to separate the speech mixture from background noise, followed by (2) a separation network trained to separate the individual speakers from the enhanced signal.

\section{Experimental results}
The WHAM! dataset allows us to evaluate multiple tasks in a controlled comparable manner.  These tasks are:
\begin{itemize}[leftmargin=6mm, topsep=0pt]
    \setlength{\itemsep}{0pt}
    \setlength{\parsep}{3pt}
    \setlength{\parskip}{3pt}
        \item \textbf{enhance-single}: from a mixture of only the first WSJ0 speaker and noise, recover the signal from the first speaker (typical speech enhancement scenario);
        \item \textbf{enhance-both}: from a mixture of two speakers and noise, recover the mixture of two speakers (rather than the separated speech signals);
        \item \textbf{separate-clean}: from a mixture of two speakers, recover the signals from each speaker (equivalent to wsj0-2mix);
        \item \textbf{separate-noisy}: from mixtures of two speakers in noise, recover the signals from each speaker.
    \vspace{3pt}
\end{itemize}

Unless otherwise stated, all neural network architectures reported in this section are re-trained for each task and follow the chimera++ architecture from \cite{Wang2018Interspeech09}, containing four BLSTM layers with 600 units in each direction, followed by dense output layers for both the mask inference and deep clustering heads. A dropout of $0.3$ is applied on the output of each BLSTM layer except the last one. The networks are trained on 400-frame segments using the Adam algorithm. The window length is 32~ms and the hop size is 8~ms. The square root Hann window is employed as the analysis window, and the synthesis window is designed to achieve perfect reconstruction after overlap-add.  Most published results we are aware of using the wsj0-2mix dataset use the 8 kHz min version of the dataset.  While this is understandable since the min version contains a higher percentage of purely overlapped speech and the compute requirements for 8 kHz models are lower, for WHAM! we present results on both the 8 kHz min version to compare with existing literature, and the 16 kHz max version to see how approaches scale up to more realistic scenarios.  We evaluate performance using the scale-invariant signal-to-distortion ratio (SI-SDR) between the target speech and the estimate \cite{LeRoux2018SISDR}.

\subsection{Oracle results}

To assess the difficulty of the different WHAM! tasks, we perform evaluation under oracle conditions (i.e., the masks are obtained via the ground truth reference signals).  Table~\ref{table:oracle} compares oracle performance using three mask types: ideal ratio mask (IRM: $a^{\mathrm{IRM}} = |s|\big/ (|s| + |n|) $), ideal binary mask (IBM: $a^{\mathrm{IBM}} = \delta(|s| > |n|)$), and phase sensitive filter (PSF: $a^{\mathrm{PSF}} = \cos(\theta) \frac{|s|}{|x|} x $), with $s$ a target, $n$ an interference, $x$ their mixture, and $\theta$ the phase angle between $s$ and $x$.  While the noisy SI-SDR is lower for 16 kHz max compared to 8 kHz min, oracle performance is similar at both sampling rates for all of the tasks.  We also note that SI-SDR improvement from noisy in the enhance-both case is about 2 dB lower than in the enhance-single case, %
suggesting that removing noise from mixtures of multiple speakers is harder than removing noise from one speaker, even without trying to separate the speakers.

\begin{table}[htbp]
    \footnotesize
    \caption{SI-SDR [dB] oracle performance on WHAM! tasks}\vspace{-0.6cm}
    \begin{center}
      \label{table:oracle}
      \begin{tabular}{lccccc}\hline
        Task & Dataset & Noisy & IRM & IBM & PSF \\ \hline \hline
        \multirow{2}{*}{enhance-single} & \phantom{1}8 kHz min & -0.9 & 11.0 & 11.6 & 14.7  \\
         & 16 kHz max & -2.9 & 11.0 & 11.6 & 14.8  \\ \hline
        \multirow{2}{*}{enhance-both} & \phantom{1}8 kHz min & \phantom{-}1.2 & 10.9 & 11.4 & 14.6 \\ 
         & 16 kHz max & -0.7 & 10.8 & 11.4 & 14.5  \\ \hline
        \multirow{2}{*}{separate-clean} & \phantom{1}8 kHz min & \phantom{-}0.0 & 12.7 & 13.5 & 16.4 \\ 
         & 16 kHz max & \phantom{-}0.0 & 13.4 & 14.2  & 17.1\\ \hline
        \multirow{2}{*}{separate-noisy} & \phantom{1}8 kHz min & -4.5 & \phantom{1}8.3 & \phantom{1}8.9  & 12.3 \\ 
         & 16 kHz max & -5.8 & \phantom{1}8.5 & \phantom{1}9.1  & 12.5 \\ \hline
      \end{tabular}\vspace{-.7cm}
    \end{center}
  \end{table}

\subsection{Model comparisons}
Table~\ref{table:compare} presents results for the chimera++ architecture on the WHAM! dataset.  For the enhancement tasks, we use a weight of $\alpha=0$ in~\eqref{eq:chimera} as deep clustering did not improve performance, while for separation tasks we use $\alpha=0.975$ and $\mathcal{L}_{\text{DC},\text{W}}$ from~\eqref{eq:dc_white} as the deep clustering objective.  For both enhancement tasks, we see a larger SI-SDR improvement in the 16 kHz max case than with 8 kHz min, likely because it is easy to enhance in regions where noise and speech do not overlap.  However, we notice a rather large drop in performance between 8 kHz and 16 kHz for separate-clean, as well as a more moderate drop for separate-noisy. %

 \begin{table}[tbp]
    \footnotesize
    \caption{SI-SDR [dB] performance comparison of chimera++ networks on WHAM! tasks, where $\Delta$ indicates  improvement.}\vspace{-0.6cm}
    \begin{center}
      \label{table:compare}
      \begin{tabular}{lcccc}\hline
        Task & Dataset & Noisy & Output & $\Delta$ \\ \hline \hline
        \multirow{2}{*}{enhance-single} & \phantom{1}8 kHz min & -0.9 & 10.2 & 11.1  \\
         & 16 kHz max & -2.9 & 10.0 & 12.9  \\ \hline
        \multirow{2}{*}{enhance-both} & \phantom{1}8 kHz min & \phantom{-}1.2 & \phantom{1}9.4 & \phantom{1}8.2  \\ 
         & 16 kHz max & -0.7 & \phantom{1}9.3 & 10.0  \\ \hline
        \multirow{2}{*}{separate-clean} & \phantom{1}8 kHz min & \phantom{-}0.0 & 11.0 & 11.0  \\ 
         & 16 kHz max & \phantom{-}0.0 & \phantom{1}9.6 & \phantom{1}9.6  \\ \hline
        \multirow{2}{*}{separate-noisy} & \phantom{1}8 kHz min & -4.5 & \phantom{1}5.4 & \phantom{1}9.9  \\ 
        & 16 kHz max & -5.8 & \phantom{1}4.4 & 10.2  \\ \hline
      \end{tabular}\vspace{-.5cm}
    \end{center}
  \end{table}

\begin{figure}[h]
	\centering
		\includegraphics[width=.49\columnwidth]{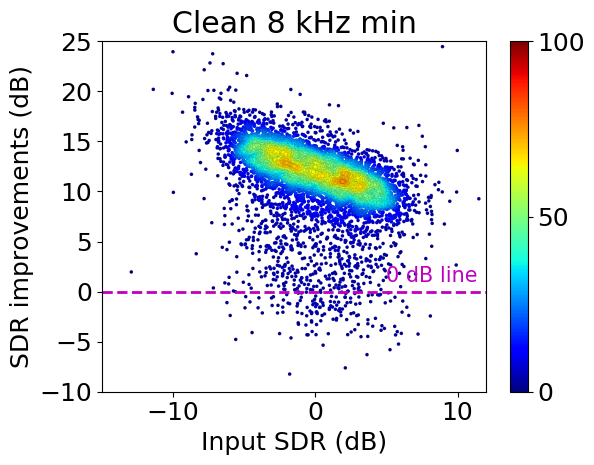}
		\includegraphics[width=.49\columnwidth]{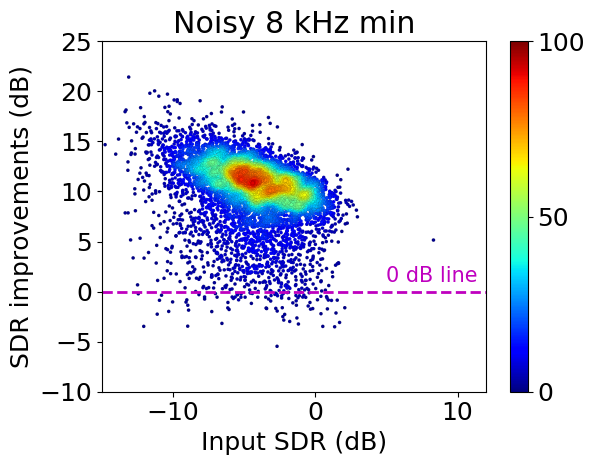}
		\includegraphics[width=.49\columnwidth]{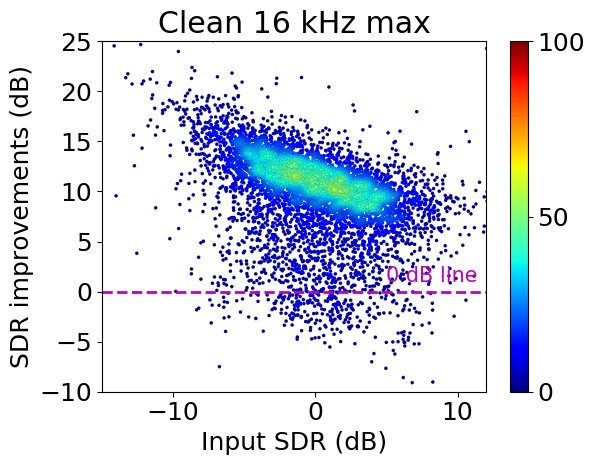}
		\includegraphics[width=.49\columnwidth]{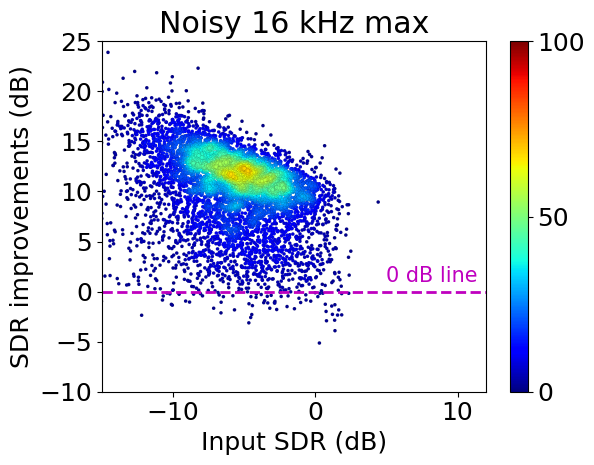}
	\caption{SI-SDR scatter plots comparing chimera++ performance over different datasets.} 
	\label{fig:scatter}
\end{figure}

To further investigate these differences, we created 2-D histogram-like scatter plots for the separate-clean and separate-noisy cases, shown in Fig.~\ref{fig:scatter}.  We see that in all cases most utterances cluster around 10 dB improvement in SI-SDR.  For the separate-clean cases (left side of Fig.~\ref{fig:scatter}), the amount of SDR improvement is much higher when the input (noisy) SDR is lower, but this improvement for very noisy speech signals is less pronounced in the noisy cases (right side of Fig.~\ref{fig:scatter}).  This suggests that improving the quality of relatively quiet speakers is more difficult in the presence of background noise.  We also hypothesize that some of the performance difference between the 16 kHz and 8 kHz case is that frame-level permutation mistakes as discussed in~\cite{aihara2019teacher} (where the speaker being tracked by the network changes mid-utterance) are more likely in the 16 kHz max case due to longer regions with only a single active speaker.

A comparison of the different deep clustering modifications discussed in Section~\ref{sec:objectives} for speech separation in noise are provided in Table~\ref{table:chimera_compare}.  The best performance is obtained with three deep clustering sources (treating noise as a source) and using the unmodified whitened k-means objective $\mathcal{L}_{\text{DC},\text{W}}$.  Handling noise by using only two deep clustering sources and removing bins without speech via weighting, or using the $\mathcal{L}_{\text{DC},\text{N}}$ objective from~\eqref{eq:dc_noisy} do not perform as well.  Table~\ref{table:chimera_compare} also provides results for the approach consisting of two different networks, the first removing the noise, and the second separating the speech signals. Without finetuning, the combined system does not perform as well as the best performing chimera++ approaches. However, if we finetune the separate-clean model on the outputs of the trained enhance-both model, the combined system outperforms all the jointly trained chimera++ approaches.  While this method is more computationally expensive, it may be useful for systems with a pre-existing enhancement algorithm.

\begin{table}[h]
    \footnotesize
    \caption{SI-SDR [dB] improvement comparison of different chimera++ objectives for noisy separation on 8 kHz min}\vspace{-0.5cm}
    \begin{center}
      \label{table:chimera_compare}
      \begin{tabular}{lcc}\hline
        DPCL Objective & DPCL Sources & $\Delta$ SI-SDR \\ \hline \hline
        n/a (mask inference) & - & \phantom{1}8.5  \\
        $\mathcal{L}_{\text{DC},\text{C}}$ & 3 & \phantom{1}9.6  \\
        $\mathcal{L}_{\text{DC},\text{N}}$ & 3 & \phantom{1}9.6  \\
        $\mathcal{L}_{\text{DC},\text{W}}$ & 3 & {\bf \phantom{1}9.9}  \\
        $\mathcal{L}_{\text{DC},\text{W}}$, 0 weight on noise bins & 2 & \phantom{1}8.4  \\ \hline
        enh-both + sep-clean & 2 & \phantom{1}9.0  \\
        enh-both + sep-clean-finetune & 2 & {\bf 10.3}  \\ \hline
      \end{tabular}
    \end{center}\vspace{-0.6cm}
  \end{table}

\subsection{Other benchmarks}
\label{sec:benchmarks}
In addition to chimera++, we also evaluated our implementation of the original TasNet algorithm~\cite{Luo2018}, using an input filter-size of 80 samples with a stride of 40 samples and 500 bases (for the STFT-like convolution/deconvolution layers), the same BLSTM stack used for chimera++, and the SI-SNR objective proposed by the authors.  We also implemented a fully convolutional model inspired by~\cite{LeaFVRH16} taking magnitude spectrograms as input, treating frequencies as input/output channels, and consisting of seven blocks of 1-d dilated convolutions followed by 1x1 convolutions with residual connections and batch norm between all layers.  Table~\ref{table:clean_v_noisy} compares these benchmarks with chimera++ on the separation tasks.  %
We see that while TasNet significantly outperformed chimera++ in the clean case, it did not perform as well under noisy conditions.  We suspect this is because learning directly from waveforms is more difficult with noisy signals.  Our 1-d convolutional model is related to (but slightly simpler than) the dilated convolution models in ~\cite{Luo2018TasNet09arXiv, shi2019furcanext}.  Like chimera++, it operates directly on the spectrogram, and while performance in terms of SI-SDR is not as high as chimera++, it trains much faster and uses fewer parameters.
  
\begin{table}[h]
    \footnotesize
    \caption{SI-SDR [dB] comparison of our implementations of other benchmark networks on the WHAM! separate-clean and separate-noisy tasks}\vspace{-0.5cm}
    \begin{center}
      \label{table:clean_v_noisy}
      \scalebox{0.96}{
      \begin{tabular}{lc|cc|cc}\hline
         & & \multicolumn{2}{c|}{separate-clean}  & \multicolumn{2}{c}{separate-noisy} \\
        Model & Dataset & Output & $\Delta$ & Output & $\Delta$ \\ \hline \hline
        chimera++ & \phantom{1}8 kHz min & 11.0 & 11.0 & 5.4 & \phantom{1}9.9  \\
        TasNet-BLSTM  & \phantom{1}8 kHz min & 12.5 & 12.5 & 5.3 & \phantom{1}9.8  \\ \hline
        chimera++ & 16 kHz max & \phantom{1}9.6 & \phantom{1}9.6 & 4.4 & 10.2  \\
        1-d conv.  & 16 kHz max & \phantom{1}6.9 & \phantom{1}6.9 & 3.0 & \phantom{1}8.8  \\ \hline
      \end{tabular}}\vspace{-.6cm}
    \end{center}
  \end{table}

\section{Conclusion}
To help move the rapidly advancing speech separation field towards more realistic scenarios, we introduced the WHAM! dataset for evaluation of speaker-independent separation in noisy environments, and used it to benchmark several speech enhancement and speech separation approaches. Initial results show that T-F based separation approaches still perform effectively in the presence of noise.
Future work includes evaluating stereo approaches for noisy speech separation, evaluating robustness to reverberation plus noise, and further exploration of the convolutional models discussed in Section~\ref{sec:benchmarks}.

\vfill\pagebreak
\balance

\bibliographystyle{IEEEtran}

\bibliography{refs}

\end{document}